\documentclass[aps,pr,twocolumn,groupedaddress,notitlepage,
floatfix,superscriptaddress]{revtex4-2}

\pdfoutput=1 
\usepackage{graphicx,graphics,epsfig,subfigure,times,bm,bbm,amssymb,amsmath,amsthm,mathrsfs,MnSymbol} \usepackage{gensymb} \usepackage{amsfonts} \usepackage{float} \usepackage[matrix,frame,arrow]{xypic} \usepackage[pdfstartview=FitH]{hyperref} 
\usepackage{times}
\usepackage{float}
\usepackage{graphics}
\usepackage[T1]{fontenc}
\usepackage{CJKutf8}

\usepackage{braket}  
\usepackage{enumerate} \usepackage[normalem]{ulem}
\usepackage[usenames,dvipsnames]{xcolor} \usepackage{multirow} \usepackage{mathtools}
\usepackage{bbm} \usepackage{titletoc} 

\definecolor{orange}{rgb}{1,0.5,0} 
\hypersetup{ colorlinks=true,       
	linkcolor=red,          
	citecolor=blue,        
	filecolor=magenta,      
	urlcolor=blue,           
	runcolor=cyan }

\begin{document}
	\title{Identifying  Entanglement Phases with Bipartite Projected Ensembles}
	
	\author{Zi-Yong~Ge}
\affiliation{Quantum Information Physics Theory Research Team, Center for Quantum Computing, RIKEN, Wako-shi, Saitama 351-0198, Japan}

	\author{Franco Nori}
\affiliation{Quantum Information Physics Theory Research Team, Center for Quantum Computing, RIKEN, Wako-shi, Saitama 351-0198, Japan}
\affiliation{Department of Physics, University of Michigan, Ann Arbor, Michigan 48109-1040, USA}

\begin{abstract}
We introduce bipartite projected ensembles (BPEs) for quantum many-body wave functions, 
which consist of pure states supported on two local subsystems, 
with each state associated with the outcome of a projective measurement of the complementary subsystem in a fixed local basis.
We demonstrate that the corresponding ensemble-averaged entanglements (EAEs) between two subsystems can effectively identify entanglement phases.
In volume-law entangled states, EAE converges to a nonzero value with increasing distance between subsystems. 
For critical systems,  EAE exhibits power-law decay, and it decays exponentially for area-law systems. 
Thus, entanglement phase transitions can be viewed as a disordered-ordered phase transition. 
We also apply BPE and EAE to measured random Clifford circuits to probe measurement-induced phase transitions. 
We show that EAE serves not only as a witness to phase transitions, 
but also unveils additional critical phenomena properties, including dynamical scaling and surface critical exponents. 
Our findings provide an alternative approach to diagnosing entanglement laws, thus enhancing the understanding of entanglement phase transitions. 
Moreover, given the accessibility of measuring EAE in quantum simulators, our results hold promise for impacting quantum simulations.
\end{abstract}
	
	\maketitle
	
   \section{Introduction}

	The development of quantum information~\cite{wilde2013quantum,watrous2018theory,cheng2023noisy,zeng2019quantum} has highlighted the significance of entanglements in understanding quantum matter~\cite{zeng2019quantum,Fradkin2013,RevModPhys.80.517}. 
	For instance, in 1D local gapped quantum many-body systems, the entanglement entropies of ground states  
	has been proved to satisfy the area law~\cite{RevModPhys.82.277}.
	However, in critical systems, the entanglement entropies display a logarithmic divergence~\cite{PhysRevA.66.032110,osterloh2002scaling,PhysRevLett.90.227902,Pasquale_Calabrese_2004}. 
    In addition to low-energy physics, entanglements can also shed light on  high-energy excited states and non-equilibrium  systems~\cite{RevModPhys.83.863}. 
    Specifically, the singular changes of entanglement scaling 
    of pure quantum many-body states can be identified by entanglement phase transitions.
    One typical example is the phase transition between thermalization and many-body localization,
    where high-energy excited states of the former system obey a volume-law, 
    while the latter one has area-law entangled excited states~\cite{Basko2006,PhysRevB.82.174411,PhysRevLett.113.107204,PhysRevB.90.174202,PhysRevB.93.041424,Rahul2015,RevModPhys.91.021001}. 
	Another novel family of entanglement phase transitions, known as measurement-induced phase transitions (MIPTs)~\cite{PhysRevB.98.205136,PhysRevX.9.031009,PhysRevB.99.224307,PhysRevB.100.064204,PhysRevB.100.134306,
	PhysRevB.101.104301,PhysRevB.101.104302,PhysRevLett.125.030505,PhysRevX.10.041020,PhysRevLett.125.070606,
	PhysRevB.101.104302,PhysRevB.101.060301,PhysRevB.102.014315,PhysRevLett.125.210602,PhysRevX.11.011030,PhysRevLett.126.060501,
	PhysRevLett.126.170602,PhysRevB.103.224210,PhysRevResearch.3.023200,PhysRevLett.128.010604,PhysRevLett.128.010603,PhysRevLett.128.010605,
	10.21468/SciPostPhysCore.5.2.023,PhysRevB.105.L241114,PhysRevX.12.041002,PhysRevLett.131.020401,PhysRevLett.132.110403,PhysRevLett.132.030401,PRXQuantum.5.030329,arXiv:2402.13271,PhysRevLett.132.163401}, 
	has been discovered in monitored quantum many-body systems,
	and is currently under experimental investigation~\cite{noel2022measurement,koh2023measurement,google2023measurement} in various quantum simulators~\cite{buluta2009quantum,trabesinger2012quantum,RevModPhys.86.153}. 
	The different measurement-induced phases manifest in different behaviors of entanglements in steady states. 
	For small rates of measurements, the system resides in an entangling phase with volume-law entanglements of steady states. 
	For high measurement rates, it is disentangling phase with area-law entanglements. 
	At the critical point, the steady states generally exhibit logarithmic entanglement entropies, satisfying the sub-volume law.

	Entanglement dynamics in some random circuits can be mapped to classical statistical mechanics~\cite{PhysRevB.101.104301,PhysRevB.101.104302}, 
where the entanglement entropies correspond to free energies and the entanglement phase transitions can be mapped to ordered-disordered phase transitions.
However, in conventional phase transitions, we generally need to define the corresponding correlation functions to identify short- and long-range order.
Moreover, correlation functions can guide us to a universal scaling law, which is significant for understanding the phase transition and critical phenomenon.
For entanglement phases, one potential candidate is quantum mutual information~\cite{PhysRevLett.100.070502}, 
which characterizes the quantum correlation of two subsystems.
However,  quantum mutual information between two local subsystems is akin to a connected correlation function, 
which tends to zero for both volume- and area-law entangled states~\cite{PhysRevB.100.134306},
so it cannot distinguish between ordered and disordered phases.
Thus, an open question arises:  whether we can define a correlation function to further understand entanglement phase transitions.

  In this paper, we introduce a type of correlation function with bipartite projected ensembles (BPEs) to identify  entanglement phases. 
Here, BPE is a set of pure states supported on two local subsystems, 
where each element represents an output state obtained after a projective measurement of the complementary subsystem in a fixed local basis. 
We also define ensemble-averaged entanglements (EAEs) between these two subsystems, which can be regarded as a type of correlation function. 
We show that the scaling of EAEs can identify the entanglement laws of quantum many-body wave functions. 
In the case of volume-law entanglements, EAE converges to a nonzero value for large distances between two subsystems. 
In a critical system, EAE follows power-law decay, and it exhibits exponential decay in an area-law system. 
We then apply BPE and EAE to investigate the MIPT in a random Clifford circuit with local projective measurements,
and our results show that \textit{EAE can effectively witness MIPT}.
Remarkably, we demonstrate that \textit{BPE can reveal the universal scaling of the dynamics and surface critical exponents},
which are significant for understanding critical properties of MIPT.
We also  propose an experimental protocol  for measuring EAE and estimate the corresponding complexity,
where our results show that  measuring EAE is easier than measuring  entanglement entropies.

       \begin{figure*}[t] \includegraphics[width=0.95\textwidth]{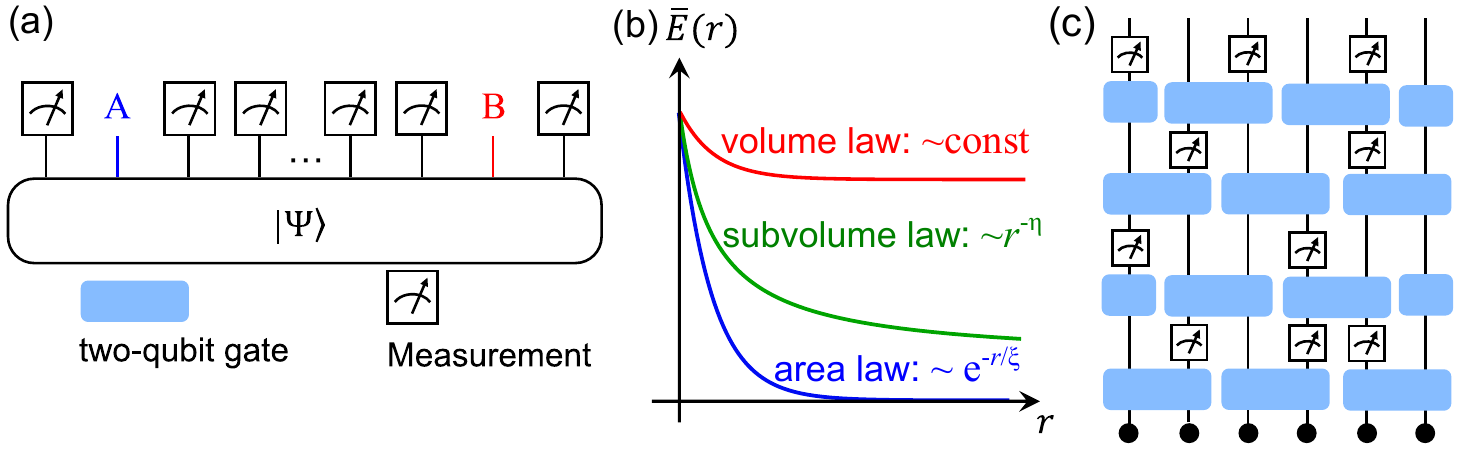}
    	\caption{Setup. (a) The diagram of BPE. For a many-qubit system with a pure state$\ket{\Psi}$, 
    		the subsystem $R$ is measured in a fixed local basis. 
    		Then, the remaining unmeasured subsystems $A$ and $B$ are in a pure state, which depends on the measurement outcome on $R$. 
    		(b) The scaling of EAE for different entanglement laws. 
    		(c) The structure of the unitary-measurement hybrid circuit. 
    		The local two-qubit gates are drawn from the uniformly sampled Clifford group, 
    		and the random projected measurement is onto the $z$-component with a probability $p$.}
    	\label{fig_1}
    \end{figure*} 
	
	\section{Setting up the bipartite projected ensembles}
	
	Here, BPE is generalized from projective ensembles in Refs.~\cite{choi2023preparing,PhysRevLett.128.060601,PRXQuantum.4.010311}.
    Without loss of generality, we consider a many-qubit system, 
    where the system is divided into three parts $A$, $B$ and $R$, with the number of qubits $N_A$, $N_B$ and $N_R$.
    The local bases of each qubit are labeled by $\ket{z}$ with $z=0,1$, and $\ket{\Psi}$ denotes the wave function of the full system.
    Here, the BPE of $A$ and $B$ is generated by performing projective measurements on all qubits in $R$ with a local basis $\{\ket{z_R}\}$,
    where $\ket{z_R}\in \{0,1\}^{N_R}$ are the measurement results of $R$ with probability $p(z_R)$, see Fig.~\ref{fig_1}(a).
    Thus, the wave function $\ket{\Psi}$ can be rewritten as 
        \begin{align} 
    	\ket{\Psi} = \sum _{z_R}\sqrt{p(z_R)}\ket{\Psi_{AB}(z_R)}\otimes \ket{z_R}
    \end{align}	
    with
    \begin{align} 
    &	p(z_R)=\bra{\Psi}( \mathbb{I}_{AB}\otimes\ket{z_R}\bra{z_R})\ket{\Psi},\\
    &	\ket{\Psi_{AB}(z_R)}=(\mathbb{I}_{AB}\otimes\bra{z_R})\ket{\Psi}/\sqrt{p(z_R)},
    \end{align}	
    where $ \mathbb{I}_{AB}$ is the identity of subsystems $A$ and $B$.
    Thus, the BPE for subsystems $A$ and $B$ can be defined as 
     \begin{align} 
  	\mathcal{E}_{\Psi, AB} := \{p(z_R),\ket{\Psi_{AB}(z_R)}\}.
   \end{align}

   To measure the correlation between subsystems $A$ and $B$, we also define the corresponding EAE as
   \begin{align} \label{eae}
   \bar E(A:B) := \sum_{z_R} p(z_R) S_A(z_R)=\sum_{z_R} p(z_R) S_B(z_R),
  \end{align}	
   where $S_{A/B}(z_R)$ is the von Neuman entropy of subsystem $A/B$ with respect to the state $\ket{\Psi_{AB}(z_R)}$.
   Here, both the BPE and EAE depend on the measurement bases.
   The maximum value of $\bar E(A:B)$ over the allowed measurement bases is the localizable entanglement~\cite{PhysRevLett.92.027901,PhysRevA.71.042306},
   which has been used to study ground-state quantum phase transitions and multipartite entanglements.
   
   In this work, we primarily utilize BPE and EAE to study entanglement phases, where we only consider a fixed basis.
   Our setting is suitable for numerical simulations and quantum simulators~\cite{choi2023preparing}.
   In the following discussion, for simplicity, we only consider $N_A=N_B=1$, and $r$ labels the distance between $A$ and $B$, 
   i.e., $\bar E(A:B) =\bar E(r)$.

   \begin{figure*}[t] \includegraphics[width=1\textwidth]{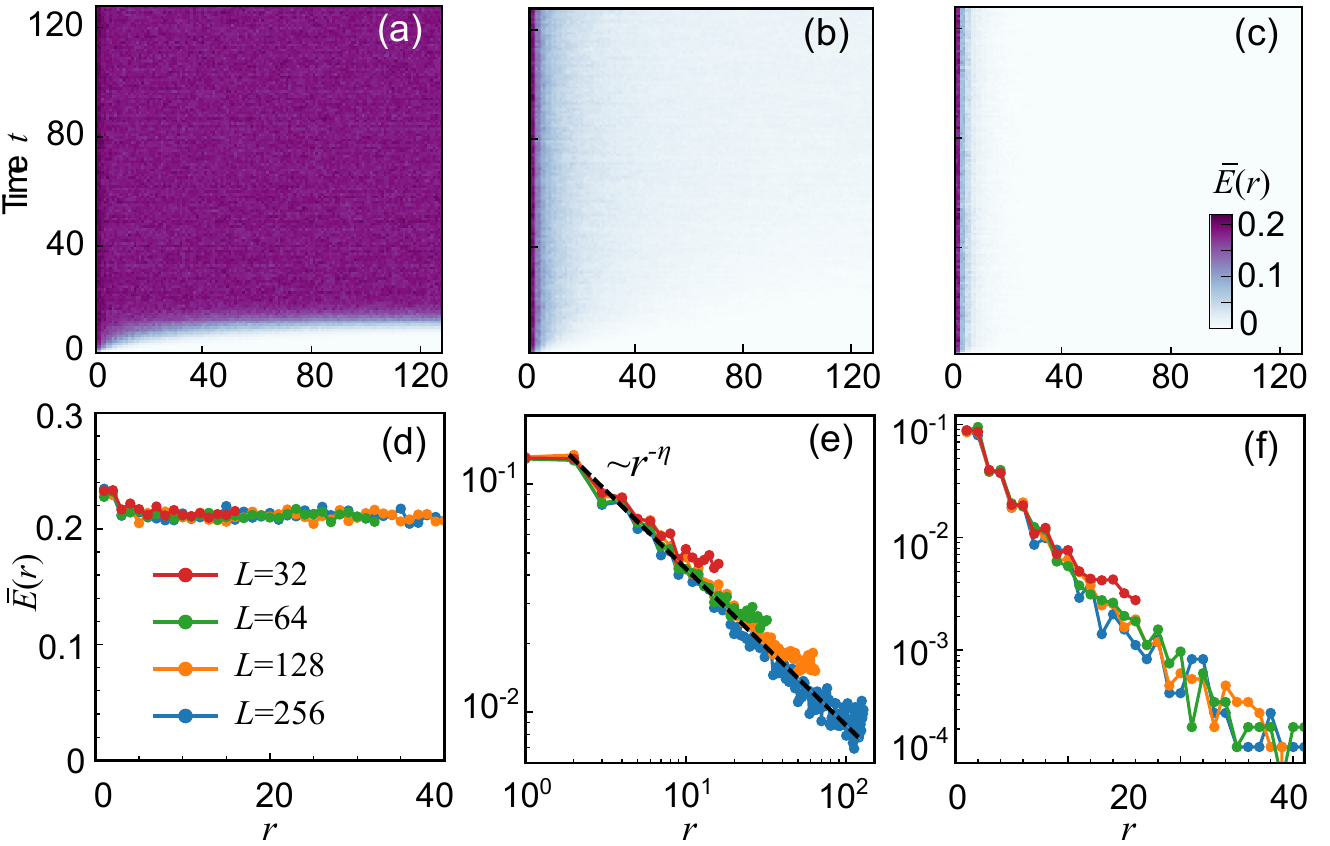}
	\caption{The results of EAE for random Clifford circuits.
	The dynamics of $\bar E(r)$ for (a) $p=0.05$, (b) $p=0.16$, and (c) $p=0.25$. The system size is $L=256$.
    The scaling of $\bar E(r)$ at $t=2L$ for (d) $p=0.05$, (e) $p=0.16$, and (f) $p=0.25$.
    The black dashed line in (e) is for the fit: $\bar E(r)\sim r^{-\eta}$, with $\eta\approx 0.71$.
    The sampling times are 10,000 for $L=32,64,128$, and 5,000 for $L=256$.}
	\label{fig_2}
   \end{figure*} 

   \begin{figure}[t] \includegraphics[width=0.48\textwidth]{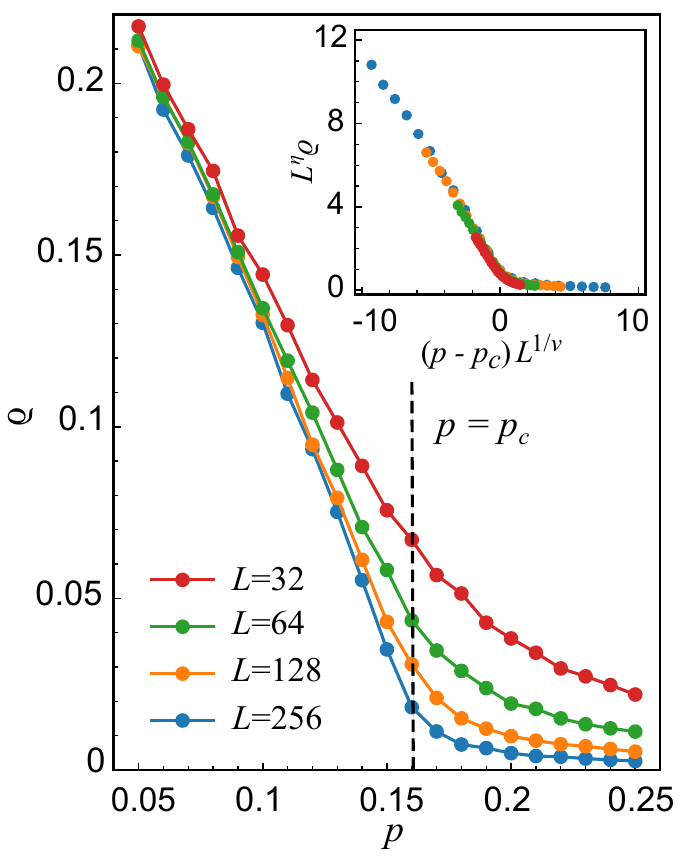}
	\caption{The integrated EAE $\varrho$ versus $p$ for different system sizes. 
		The inset is a collapse of the data, with the critical point $p_c \approx 0.16$ and exponent $\nu\approx1.24$.
		We choose periodic boundary conditions.
		The sampling times are 10,000 for $L=32,64,128$, and 5,000 for $L=256$.}
	\label{fig_3}
\end{figure} 
   
    Now we discuss the relations between EAEs and the laws of entanglement entropies.
    In the Appendix~\ref{app2}, we present phenomenological derivations of the scaling of the EAE
    in different entanglement phases with the matrix-product state representation~\cite{SCHOLLWOCK201196}:
    \begin{enumerate}
    \item[(\textit{i})] For area-law entangled states, $\bar E(r)$ exhibits an exponential decay,  
    	corresponding to a short-range correlated (disordered) phase (Appendix~\ref{app2}1).
    	In Appendix~\ref{app2}4, we also consider a valence-bond solid state as an example to illustrate it.
    	   \item[(\textit{ii})]  For volume-law entangled states, $\bar E(r)$ can converge to a nonzero value when increasing $r$
    	      \begin{align} 
    	   	\lim_{r \to \infty}\bar E(r)\sim \text{const},
    	   \end{align}	
    	corresponding to a long-range correlated (ordered) phase (Appendix~\ref{app2}2).
    	To understand this, we can consider a random state as an example (Appendix~\ref{app2}4), which is a typical volume-law entangled state 
    	describing quantum chaotic systems~\cite{PhysRevLett.71.1291,DAlessio2016}.
    	For this state, $\mathcal{E}_{\Psi, AB}$ is nearly independent of measurement bases and positions of $A$ and $B$, 
    	and tends to be a Haar ensemble leading to 
    	   \begin{align} 
    	\bar E(r) \approx (\ln 2)/2,
    	\end{align}	
    	for arbitrary $r$~\cite{choi2023preparing,PhysRevLett.128.060601,PRXQuantum.4.010311}.
    	  \item[(\textit{iii})]  For critical states, due to scaling invariance, we expect that $\bar E(r)$ exhibits a power-law decay (Appendix~\ref{app2}3).
    \end{enumerate}
    Therefore, EAE can be interpreted as an effective correlation function,
    and an entanglement phase transition can be mapped to a conventional disordered-ordered phase transition, see Fig.~\ref{fig_1}(b).


 	\section{Witness of measurement-induced phase transitions}
 	
 	\subsection{ Monitored random Clifford circuit }
 	
    Since BPE and EAE can characterize the different laws of entanglement entropies, we now apply them to investigate MIPTs.
    The monitored random Clifford circuit can be simulated efficiently on classical computers~\cite{PhysRevA.70.052328} and is a paradigmatic model to study MIPTs, which only consists of Hadamard gates, Pauli $X,\ Y,\ Z$ gates, control-not gates, and $z$-component projective measurements.
    
 	Here, we find that $\bar E(r)$ of stabilizer states can also be effectively calculated,
 	and the details are presented in Appendix~\ref{app3}.
 	Thus, for concreteness, we consider the measured stabilizer circuit model as an example to investigate BPE and EAE  across a MIPT.
 	The circuit diagram is shown in Fig.~\ref{fig_1}(c),  where each time step contains two layers of quantum gates and  measurements.
 	Each local two-qubit gate $\hat U_{j,j+1}$ is independently drawn from a uniformly sampled Clifford group,
 	and the probability of each projective measurement on the circuit is $p$.
    In addition, the measurement bases of  BPE are also of $z$-component type.
 	We start from a trivial product state $\ket{\Psi_0}=\ket{0}^{\otimes L}$, with $L$ being the system size.

 	Generally, the unitary quantum gates can produce the entanglement entropies, 
 	while the projected measurements can decrease the entanglements.
 	Thus, when tuning the measurement probability $p$, there exists an entanglement phase transition 
 	with a critical point at $p_c\approx0.16$~\cite{PhysRevB.98.205136,PhysRevB.100.134306,PhysRevLett.125.070606}.
 	When $p<p_c$, the system is in an entangling phase with extensive entanglement entropies for steady states.
 	When $p>p_c$, it is in a disentangling phase with area-law entanglement entropies.
 	At the critical point, the entanglement entropies of the steady states hold a sub-volume law.

 	 	   \begin{figure*}[t] \includegraphics[width=0.95\textwidth]{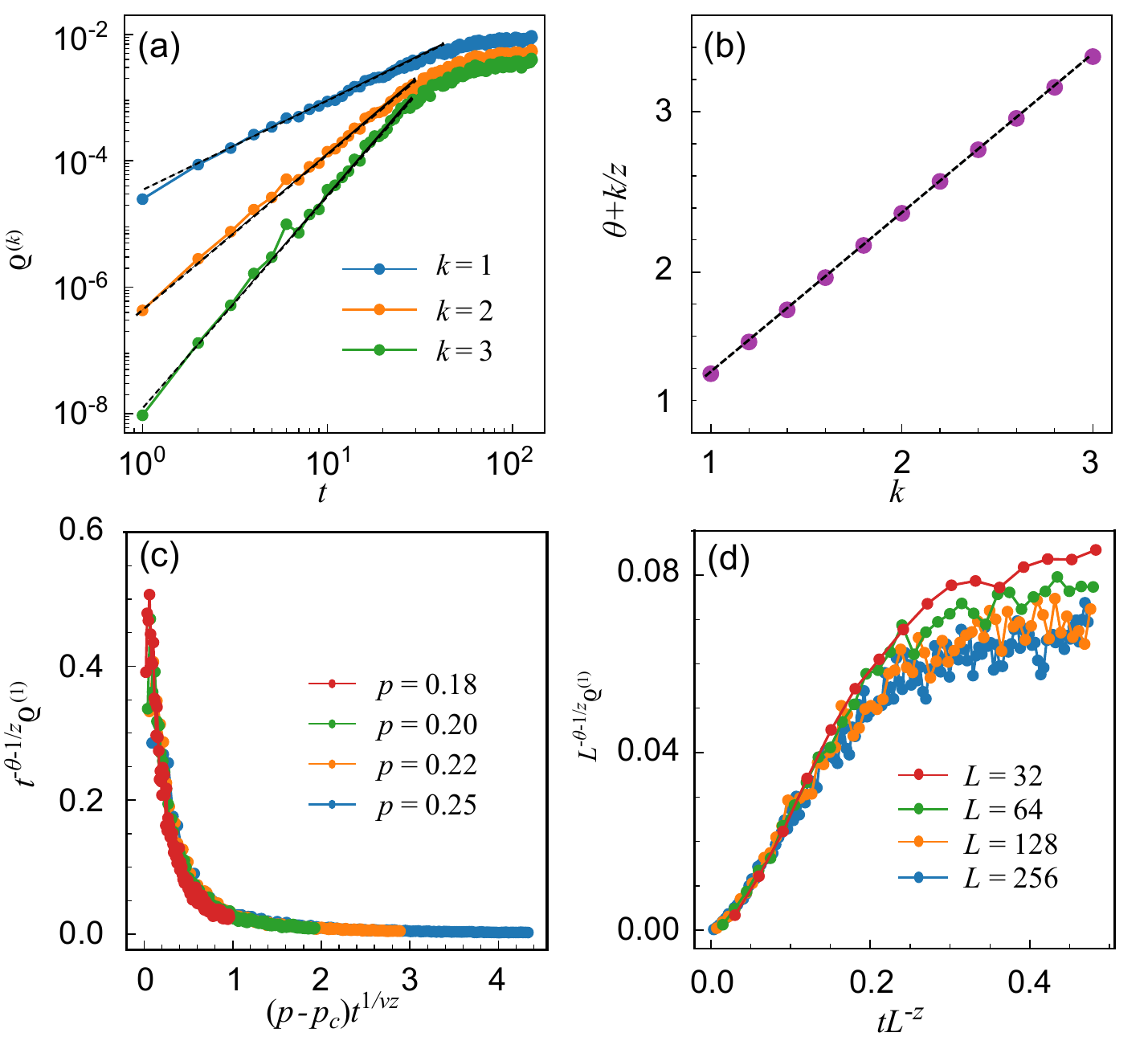}
 		\caption{Critical dynamic scaling. (a) The dynamics of $\varrho^{(k)}$ at the critical point with $L=256$.
 			The black dashed lines are linear fits.
 			(b) Scaling of $\varrho^{(k)}$ versus $k$. The black dashed line is a linear fit with $\theta \approx0.38$ and $z \approx1.01$.
 			(c) The dynamics of $\varrho^{(1)}$ in the disentangling phase for different $p$ with a fixed system size $L=256$.
 			(d) Finite-size scaling of $\varrho^{(1)}(t)$ at the critical point.}
 		\label{fig_4}
 	\end{figure*} 

   \subsection{ Phase transitions }
 	Now we numerically study the behaviors of BPE and EAE across a MIPT.
 	In Figs.~\ref{fig_2}(a-c), we present the dynamics of EAEs for different $p$.
 	We can find that, for small $p$, the correlation can spread to the whole system,
 	indicating extensive entanglement after a long-time evolution.
 	However, for large $p$, there only exist short-range correlations during the dynamics, 
 	showing area-law entanglements.
 	To further demonstrate this picture, we also present the results of $\bar E(r)$ for steady states ($t=2L$). 
 	When $p<p_c$, $\bar E(r)$ can converge to a nonzero value for large distances, 
 	i.e., $\bar E(r\to \infty)\sim \text{const}$, see Fig.~\ref{fig_2}(d).
 	When $p=0.16$ (critical point), EAE holds a perfect power-law decay [Fig.~\ref{fig_2}(e)]
 	 	\begin{align} 
 	\bar E(r)\sim r^{-\eta},
 	\end{align}	
 	with 
 	\begin{align} 
 	\eta \approx0.71.
 	\end{align}	
    When $p>p_c$, $\bar E(r)$ tends to exhibit an exponential decay, see Fig.~\ref{fig_2}(f).
    Our results show that \textit{the behaviors of $E(r)$ can diagnose the different entanglement phases of the measured stabilizer circuit},
    which is consistent with Fig.~\ref{fig_1}(b).

 	To further understand MIPT with EAE, we define the integrated EAE as
 	\begin{align} 
 		\varrho := \frac{1}{L}\sum_{r=1}^{L}\bar E(r).
 	\end{align}	
   Here, $\varrho$ is in analogy with the squared magnetization in the Ising model~\cite{squareM}, 
   which is useful for identifying the magnetization phase transition.
   We present the results of $\varrho$ for steady states  versus $p$ in Fig.~\ref{fig_3}.
   Similar to  the squared magnetization, in the ordered phase (small $p$),  $\varrho$ can be nonzero in the thermodynamic limit,
   while it tends to zero for the disordered phase (large $p$).
   Moreover, $\varrho$ is also expected to satisfy the scaling ansatz (Appendix~\ref{app4}1)
   \begin{align} 
   	\varrho = L^{-\eta}F[(p-p_c)L^{1/\nu}].
   \end{align}	
 	By data collapse,  we can obtain the optimal correlation length critical exponent  [see the inset in Fig.~\ref{fig_3}]
 	   \begin{align} 
 		\nu\approx1.24,
 	\end{align}	
 	which is consistent with the results in Refs.~\cite{PhysRevB.98.205136,PhysRevB.100.134306,PhysRevLett.125.070606}.
 	Therefore, by introducing EAEs, MIPTs can indeed be described with the language of conventional disordered-ordered phase transitions.
 	 In  Appendix~\ref{app1}, we demonstrate that BPE and EAE can also  identify  MIPTs in measured Haar random circuits.

    \subsection{Critical dynamic scaling}
 	We also study the critical dynamic scaling of EAE in this system.
 	Generally, the length scales in the spatial ($\xi$) and the temporal ($\tau$) directions satisfy the relation: $\tau\sim\xi^z$, 
 	where $z$ is the dynamical critical exponent.
 	To identify the universal dynamics of EAE, we first need to obtain $z$.
 	In previous works~\cite{PhysRevB.98.205136,PhysRevLett.125.070606}, $z$ is usually taken as $1$.
 	However, there has not been any reliable analytical or numerical results to verify $z=1$ in this system~\cite{PhysRevB.98.205136,PhysRevLett.125.070606}.
 	Here, we demonstrate that EAE can be used to obtain $z$.
 	In analogy to conventional critical non-equilibrium systems, the dynamics of EAE is expected to satisfy the scaling ansatz~\cite{Altland2010}
 	 \begin{align} \label{Ert}
 		\bar E(r,t,p) = t^{\theta-d/z} g[r/t^{1/z},(p-p_c)t^{1/\nu z}],
 	 \end{align}	
     where $d=1$ is the spatial dimension, and $\theta$ is a universal critical exponent.
    Here, we can also define the $k$-moment of EAE as 
        \begin{align} \label{rhokt}
    	\varrho^{(k)}:= \frac{1}{L^{k+1}}\sum_{r} r^k \bar E(r),
    \end{align}
    where $\varrho = \varrho^{(k=0)}$.
    According to Eq.~(\ref{Ert}), we can obtain the dynamics of $\varrho^{(k)}$ by integrating $r$ as (Appendix~\ref{app4}2)
    \begin{align} \label{rhokt_s}
    		\varrho^{(k)}(t,p) = t^{\theta+k/z} G_k[(p-p_c)t^{1/\nu z}].
     \end{align}
    Thus, in the thermodynamic limit $L\rightarrow \infty$ and at the critical point,  
    the early-time dynamics of $\varrho^{(k)}$ satisfies (Appendix~\ref{app4}2)
     \begin{align} 
    	\varrho^{(k)}(t,p_c) = t^{\theta+k/z} G_k(0)\sim t^{\theta+k/z}.
    \end{align}	
     In Fig.~\ref{fig_4}(a), we present the dynamics of $\varrho^{(k)}$ at the critical point for $L=256$,
     where $\varrho^{(k)}$ indeed exhibits  power-law increasing for early times.
     By linear fitting, we can obtain the optimal critical exponents $z\approx1.01$, 
     and $\theta\approx0.38$, see Fig.~\ref{fig_4}(b).

    Now we can further verify the universal dynamics of $\varrho^{(k)}$.
    In Fig.~\ref{fig_4}(c), we plot the dynamics of $\varrho^{(1)}$ in the disordered phase.
    The perfect data collapse is consistent with the scaling ansatz in Eq.~(\ref{rhokt_s}).
    In addition, at the critical point, the early-time dynamics of $\varrho^{(k)}$ can be generalized to the finite-size scaling form
     \begin{align} \label{rhoktpc}
    	 \varrho^{(k)}(t) = L^{\theta+k/z} G(tL^{-z}).
    \end{align}	
     In Fig.~\ref{fig_4}(d), we present the dynamics of $\varrho^{(1)}$ for different system sizes at the critical point,
     of which the perfect data collapse at early times is consistent with Eq.~(\ref{rhoktpc}).

     \begin{figure}[t] \includegraphics[width=0.48\textwidth]{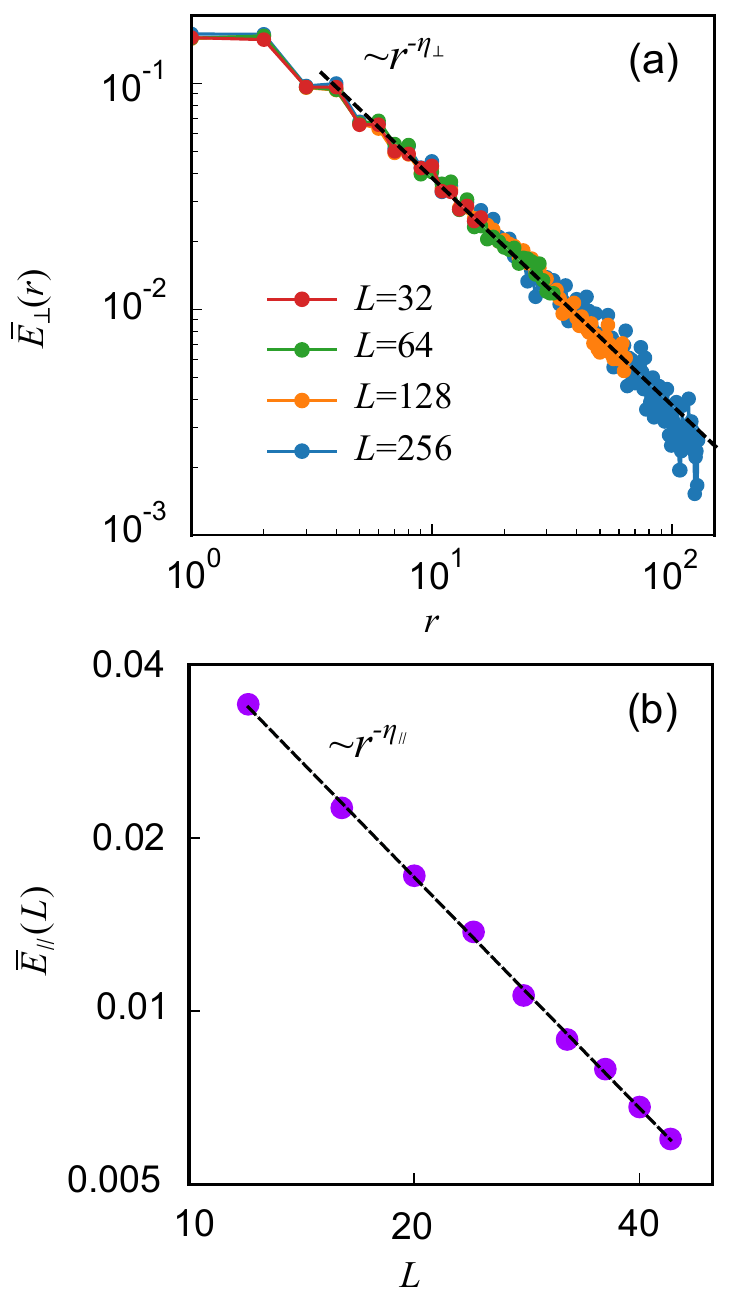}
	\caption{Surface critical phenomena. (a) The scaling of EAEs between an edge site and a bulk edge.
		(b) The scaling of EAEs between two edge sites. The black dashed lines are linear fits, with $\eta_\perp\approx1.02$ and $\eta_\parallel \approx1.34$.}
	\label{fig_5}
   \end{figure}  

    \subsection{Surface criticality}
    Now we apply EAE to identify surface criticality~\cite{CARDY1986200,CARDY1984514}, which requires open boundary conditions.
    We first consider an edge site as the subsystem $A$ and a bulk site as the subsystem $B$, respectively,
    and label the corresponding EAE as $\bar E_{\perp}(r)$.
    In Fig.~\ref{fig_5}(a), we plot $\bar E_{\perp}(r)$ of steady states at the critical point.
    The results show that $\bar E_{\perp}(r)$ can also exhibit a power-law decay: 
    \begin{align} 
    	\bar E_{\perp}(r)\sim r^{-\eta_\perp},
    \end{align}	
    with 
    \begin{align} 
    \eta_\perp\approx1.02.
    \end{align}

    Then we consider two edge sites as the subsystems $A$ and $B$, respectively, 
    and label the corresponding EAE as $\bar E_{\parallel}(L)$.
    In Fig.~\ref{fig_5}(b), we present numerical results of $\bar E_{\parallel}$ of steady states versus the system size $L$ at the critical point,
    where we can find that 
        \begin{align} 
    \bar E_{\parallel}(L)\sim L^{-\eta_\parallel},
    \end{align}	
    with 
    \begin{align} 
    	\eta_\parallel\approx1.34.
    \end{align}	
    Therefore, similar to the correlation functions in conventional phase transitions~\cite{CARDY1986200,CARDY1984514}, 
    the bulk and surface critical exponents described by EAEs in this MIPT also satisfy the relation 
    \begin{align} 
    	\eta_\perp\approx(\eta+\eta_\parallel)/2.
    \end{align}

     \section{Experimental proposal and the complexity for measuring EAE} 
     
     According to Eq.~(\ref{eae}), to measure $\bar{E}(A \! : \! B)$, it suffices to obtain the projected single-qubit density matrices of $A$, i.e., 
       \begin{align} 
     	\hat{\rho}_A (z_R) := \text{Tr}_B \ket{\Psi_{AB}(z_R)} \bra{\Psi_{AB}(z_R)}.
     \end{align}	
      Moreover, it is not necessary to access the full set of $\mathcal{E}_{\Psi, AB}$; instead, a subset of partial density matrices $\hat{\rho}_A (z_R)$ with high probabilities $p(z_R)$ is sufficient to approximate $\bar{E}(A \! :\! B)$, see also Appendix~\ref{app5}. 
      
      In experiments, a single-qubit state tomography on $A$ can be performed alongside a joint readout on $R$, and $\bar{E}(A \! : \! B)$ can be extracted by post-selection processing. Generally, the maximum $p(z_R)$ decreases exponentially with system size, implying that the complexity of probing EAE scales as $q^L$, where $q \leq 2$. In Appendix~\ref{app5}, we provide a detailed analysis of the complexity for measuring EAE, demonstrating that it is more scalable than measuring entanglement entropies~\cite{brydges2019probing} in noisy intermediate-scale quantum systems.

    \section{Conclusion}
    
        In summary, we have introduced the BPE and EAE as tools for probing entanglement phases. 
    Our findings establish that EAEs can be interpreted as a form of correlation function, whose scaling exhibits a direct correspondence with the entanglement laws of quantum many-body wave functions. 
    This insight allows us to extend concepts from conventional disordered-ordered phase transitions to the study of entanglement phase transitions,
    where volume- and area-law entangled states correspond to ordered and disordered phases, respectively.
    Using a monitored stabilizer circuit as a case study, we have demonstrated the efficacy of EAE in identifying MIPTs. 
    In particular, we have shown that EAE enables the extraction of universal scaling behavior in dynamical evolution 
    and the determination of surface critical exponents in MIPTs.
    Our results deepen the understanding of entanglement phases and offer promising avenues for advancing quantum simulation techniques.
    In addition, we anticipate the generalization of BPE and EAE to investigate other quantum many-body physics, 
    such as quantum thermalization~\cite{PhysRevA.43.2046,PhysRevE.50.888,Srednicki_1999,Rigol2008Nature,DAlessio2016} 
    and many-body localizations~\cite{Basko2006,PhysRevB.82.174411,PhysRevLett.113.107204,
    	PhysRevB.90.174202,PhysRevB.93.041424,Rahul2015,RevModPhys.91.021001}.

	\begin{acknowledgments}
      We acknowledges partial support from:
      the Japan Science and Technology Agency (JST)
      [via the CREST Quantum Frontiers program Grant No. JPMJCR24I2,
      the Quantum Leap Flagship Program (Q-LEAP), and the Moonshot R\&D Grant Number JPMJMS2061],
      and the Office of Naval Research (ONR) Global (via Grant No. N62909-23-1-2074).
	\end{acknowledgments}

\begin{appendix}

\section{Scaling of EAE}\label{app2}
In this appendix, we present phenomenological descriptions for the scaling of the EAE in different entanglement phases.
Our derivations are based on the matrix-product state (MPS) representations~\cite{SCHOLLWOCK201196}.

We write a generic many-qubit state as an MPS state with open boundary conditions 
\begin{align} 
	\ket{\Psi} = \sum_{ \bm{z} } \Gamma^{z_1}_1 \Gamma^{z_2}_2 ... \Gamma^{z_L}_L \ket{\bm{z}},
\end{align}	
where the index $\bm{z}=\{{z_1},{z_2},...,{z_L}\}$, with $z_j = 0, 1$ is the physical index.
We assume the maximum bond dimension to be $\chi$, which satisfies $2^m\leq\chi<2^{m+1}$.
Thus, the dimensions of the matrices $\Gamma^{z_j}_j$ are:
	\begin{align} \nonumber
&	(1,2),(2,4),...,(2^m,\chi),(\chi,\chi),..., \\ \nonumber
&	(\chi,\chi),(\chi,2^m),...,(4,2)(2,1).
\end{align}

	Without loss of generality, we let sites $1$ and $L$ be the subsystems $A$ and $B$, respectively.
	Thus, the projected states of $A$ and $B$ can be written  (without normalization) as
	\begin{align} 
		\ket{\Psi_{AB}(z_R)} = \sum_{ z_1,z_L } \Gamma^{z_1}_1 R(z_R) \Gamma^{z_L}_L \ket{z_1,z_L},
	\end{align}	
	where $R(z_R) $, with $z_R =\{ r_2, r_3,...,r_{L-1}\}$, is a $2\times2$ matrix and has the form
	\begin{align} 
		R(z_R)= \Gamma^{r_2}_2  \Gamma^{r_3}_3...\Gamma^{r_{L-1}}_{L-1}.
	\end{align}	
	Here, we can use the concurrence $C$ to represent the entanglement of the two-qubit state $\ket{\Psi_{AB}(z_R)}$.
	It is not difficult to verify that~\cite{PhysRevA.71.042306}
	\begin{align} 
		C(\ket{\Psi_{AB}(z_R)})= \frac{|\det [R(z_R)]|}{\text{Tr}[R(z_R) R(z_R)^\dagger]}C(\ket{\tilde{\Psi}_{AB}}),
	\end{align}	
	where $\ket{\tilde{\Psi}_{AB}} :=  \sum_{ z_1,z_L } \Gamma^{z_1}_1 \Gamma^{z_L}_L \ket{z_1,z_L}$ (without normalization).
	Here, the entanglement entropy of $\ket{\tilde{\Psi}_{AB}}$ depends on the singular values of  $\Gamma^{z_1}_1$, 
	and thus equals to the entanglement entropy of $A$ for $\ket{\Psi_{AB}}$, which is generally nonzero.
	Therefore, to calculate the scaling of the EAE, we just need to calculate the scaling of 
	\begin{align} \label{F}
		F := \frac{|\det [R(z_R)]|}{\text{Tr}[R(z_R) R(z_R)^\dagger]}.
	\end{align}	
	In the following, we will discuss the scaling of $F$ in different entanglement phases.

	\subsection{Area-law entanglement}
	We first consider area-law entangled states, which can be efficiently represented by MPS with the maximum bond dimension $\chi\sim O(1)$.
	Thus, $m\sim O(1)$.
	
	For simplicity, we let $R(z_R)$ as the product of three matrices, i.e., $R(z_R)=XWY$, where
	\begin{align}\label {lmr} \nonumber
		&X = \Gamma^{r_2}_2  \Gamma^{r_3}_3...\Gamma^{r_{m+1}}_{m+1},\\ \nonumber
		&W = \Gamma^{r_{m+2}}_{m+2}  \Gamma^{r_{m+2}}_{m+2}...\Gamma^{r_{L-m-1}}_{L-m-1},\\
		&Y = \Gamma^{r_{L-m}}_{L-m}  \Gamma^{r_{L-m+1}}_{L-m+1}...\Gamma^{r_{L-1}}_{L-1}.
	\end{align}	
	Thus, the dimensions of $X,\ W,\ Y$ are: $(2,\chi),(\chi,\chi), (\chi,2)$.
	We perform a singular-value decomposition for $W$, i.e., let $W = U \Lambda V^\dagger$,
	Here, $U$ and $V$ are both $\chi\times\chi$ unitary matrices, and $\Lambda=\text{diag}(\lambda_1,\lambda_2,...,\lambda_\chi)$,
	where we let the singular values satisfy
	\begin{align} \nonumber
		\lambda_1\geq \lambda_2\geq...\geq\lambda_\chi.
	\end{align}	
	We also define $\tilde{X} :=XU$ and $\tilde{Y} :=V^\dagger Y$, and have
	\begin{align} \nonumber
		R(z_R)=\tilde X \Lambda \tilde Y.
	\end{align}	
	We  label the elements of the matrices  $\tilde X$ and $\tilde Y$ as $x_{i,j}$ and $y_{i,j}$, respectively,
	where $|x_{i,j}| \sim  |y_{i,j}|$.
	Thus, we can obtain $ R(z_R)$ as
	\begin{align} 
		R(z_R) = \begin{bmatrix}
			\sum_{j=1}^\chi x_{1,j}\lambda_jy_{j,1}  & \sum_{j=1}^\chi x_{1,j}\lambda_jy_{j,2} \\
			\sum_{j=1}^\chi  x_{2,j}\lambda_jy_{j,1} & \sum_{j=1}^\chi  x_{2,j}\lambda_jy_{j,2}
		\end{bmatrix}.
	\end{align}	
	Therefore, we have
	\begin{align} \label{detB}
		|\det [R(z_R)]| = \sum _{i,j=1}^\chi  &\lambda_i\lambda_j |x_{1,i}x_{2,j}(y_{i,1}y_{j,2}-y_{j,1}y_{i,2})|,\\ \label{trB}\nonumber
		\text{Tr}[R(z_R) R(z_R)^\dagger] =  &\sum _{i,j=1}^\chi\lambda_i\lambda_j (x_{1,i}x^*_{1,j}   +x_{2,i}x^*_{2,j}  )\\ 
		&(y_{i,1}y^*_{j,1}  +y_{i,2}y^*_{j,2}  ).
	\end{align}

	Since $\chi\sim O(1)$ for area-law states,
	we know that the  matrix $W$ can  be described by a few singular values,
	i.e., $\lambda_j$ should decay very fast.
We first consider the non-degenerate case  $\lambda_1 \neq\lambda_2$,
		where the corresponding area-law states are generally topologically trivial and do not have long-range entanglement.
	Thus, we have 
	\begin{align} 
		{\lambda_1} \gg{\lambda_2}.
	\end{align}	
	To obtain the scaling of ${\lambda_1}/{\lambda_2}$, we can consider a  coarse-grained picture.
	We consider the subsystem $R$ (i.e., sites $m+2,\ m+3,\ m+4,\ ...,\ L-m-1$) as a cell, 
	where the corresponding tensor is $W$ and the correlation length of the new lattice is $\sim 1/L$.
	For a MPS, the correlation length is generally given by
	\begin{align} 
		\xi\sim-\frac{1}{\ln ({\lambda_2}/{\lambda_1})}\sim \frac{1}{L}.
	\end{align}	
	Thus, the maximum singular value of $M$ is exponentially larger than the second largest one, i.e.,
	\begin{align} 
		\frac{\lambda_1}{\lambda_2}\ \sim \ \exp{(\kappa L)}.
	\end{align}	
	According to Eqs.~(\ref{detB}, \ref{trB}), we have 
	\begin{align} 
		&|\det [R(z_R)]| \ \sim\  \lambda_1 \lambda_2 \\
		&\text{Tr}[R(z_R) R(z_R)^\dagger] \ \sim \  \lambda_1^2 .
	\end{align}	
	According to Eq.~(\ref{F})
	\begin{align} 
		F \ \sim \ 	\frac{\lambda_2}{\lambda_1} \ \sim \  \exp{(-\kappa L)}.
	\end{align}	
	Therefore, for these area-law entangled states, EAE is expected to exhibit an exponential decay when increasing the distance.

		Now we consider the degenerate case $\lambda_1=\lambda_2$,
		corresponding to ground states of topological systems or the GHZ state.
		As an illustrative example, we consider the GHZ state:
		\begin{align}
			\ket{\psi_{\text{GHZ}}} = \frac{1}{\sqrt{2}} \left( \ket{00\cdots 0} + \ket{11\cdots 1} \right).
		\end{align}	
		We now examine a class of homogeneous measurement bases defined as:
		\begin{align} \label{basis}
			\{\ket{\phi_1} = \sin\theta\,\ket{0} + \cos\theta\,\ket{1}, \quad \ket{\phi_2} = \cos\theta\,\ket{0} - \sin\theta\,\ket{1} \},
		\end{align}	
		with $\theta \in [-\pi/2, \pi/2]$. Each measurement outcome $\ket{z_R}$ is a product state of $\ket{\phi_1}$ and $\ket{\phi_2}$.
		The corresponding projected state of subsystems $A$ and $B$ takes the form:
		\begin{align}\nonumber
			\ket{\Psi_{AB}(z_R)} =& \mathcal{N} \big( \sin^m\theta \cos^{L-m-2}\theta \ket{00}  \\
			&+ \cos^m\theta \sin^{L-m-2}\theta \ket{11} \big),
		\end{align}	
		where $\mathcal{N}$ is a normalization factor and $m$ denotes the number of $\ket{\phi_1}$ in the outcome $\ket{z_R}$.
		Using concurrence to quantify the entanglement between $A$ and $B$, we obtain:
		\begin{align}
			C(z_R) = \mathcal{N}^2 |\sin^{L-2}\theta \cos^{L-2}\theta|.
		\end{align}	
		
		The averaged EAE over all measurement outcomes is then:
		\begin{align}
			\bar{E}_{A:B} = \sum_{m=0}^{L-2} \binom{L-2}{m} \frac{C(z_R)}{\mathcal{N}^2} = |\sin 2\theta|^{L-2}.
		\end{align}	
		Thus, $\bar{E}_{A:B}$ exhibits exponential decay for generic $\theta$, and violates it only when $\theta = \pm \pi/4$.
		Therefore, we can find that, for long-range entangled and topologically nontrivial states, 
		the EAE can violate exponential decay only for specific measurement bases.

	\subsection{Volume-law entanglement}
	For volume-law entangled states, the situation becomes completely different,
	which cannot be efficiently represented by MPS with finite maximum bond dimensions,
	i.e., the tensor cannot be truncated.
	This results from  two properties of MPS for volume-law entangled states:
	(i) Singular values of the tensor $\Gamma^{z_{k}}_{k}$ decay very slowly, i.e., they have the same order.
	(ii) The direction regarding each singular value is nearly random.
	Therefore, the matrix $R(z_R)$ should be a random matrix.
	We label two singular values of $R(z_R)$ as $s_1$ and $s_2$, which are also random.
	Thus, $s_1$ and $s_2$ are expected to satisfy 
	\begin{align} 
		\frac{s_1}{s_2}\sim O(1).
	\end{align}	
	In addition, 
	\begin{align} 
		&	|\det [R(z_R)]| = s_1s_2, \\
		&	 \text{Tr}[R(z_R) R(z_R)^\dagger]=s_1^2+s_2^2.
	\end{align}	
	Thus, 
	\begin{align} 
		\frac{|\det [R(z_R)]|}{\text{Tr}[R(z_R) R(z_R)^\dagger]} \sim O(1).
	\end{align}	
	Therefore, for volume-law entangled states, $F$ in Eq.~(\ref{F}) is finite when $L\rightarrow\infty$, 
	i.e., EAE converges to a nonzero value when increasing the distance between subsystems.

	\subsection{Critical states}
	For critical states, the entanglement entropies diverge logarithmically with respect to subsystem sizes.
	In addition, the critical states can be efficiently described by MPS with the maximum bond dimension 
		\begin{align} 
		\chi\sim L^\alpha,
	\end{align}	
	where the parameter $\alpha$ relates to the central charge.
	Here, the singular values $\lambda_j$ of the matrix $W$ defined in Eq.~(\ref{lmr})  decay slowly.
	Thus, the scaling of $|\det [R(z_R)]|$ and ${\text{Tr}[R(z_R) R(z_R)^\dagger]} $ depend on all $\lambda_j$
	and the matrices $\tilde{X}$ and $\tilde{Y}$, making it challenging to obtain the  scaling  of $F$.

	However, according to the MPS representation, 
	we know that the scaling of EAE  for critical states should be intermediate between the one of area- and volume-law states.
	In addition, due to the scaling invariance for the critical states, intuitively, EAE is expected to exhibit a power-law decay.

	\subsection{ Two solvable examples}
	Here we  present two instructive solvable examples of quantum many-body states 
	to illustrate the scalings of EAE for area- and volume-law entangled states.

	For area-law states, we consider a valence-bond solid state as an example, i.e.,
	\begin{align} 
		\ket{\psi_{\text{VBS}} }= \bigotimes^{N/2}_{j=1} \ket{\phi_{2j-1,2j}}, \ \ \ \   \ket{\phi_{2j-1,2j}}=\frac{1}{\sqrt{2}}(\ket{01}-\ket{10}).
	\end{align}	
	We choose the first site as the subsystem $A$.
	When the subsystem $B$ is site 2 (i.e., $r=1$), we can find that $\ket{\Psi_{AB}(z_R)}\}=\frac{1}{\sqrt{2}}(\ket{01}-\ket{10})$ for arbitrary $z_R$.
	Thus, the corresponding EAE is $\bar E(A \!:\! B) = \ln 2$.
	When the subsystem $B$ is not site 2 (i.e., $r>1$), 
	each element $\ket{\Psi_{AB}(z_R)}\}$ in the BPE has four possibilities:
	$ \ket{00}, \ \ket{01}, \ \ket{10}, \ \ket{11}$,
	which are all product states, leading to $\bar E(A \! : \! B) = 0$.
	Therefore, the scaling of EAE for $\ket{\psi_{\text{VBS}} }$ satisfies
	\begin{equation}
		\bar E(r) = 
		\left\{
		\begin{aligned}
			&\ln 2,\ \ \ r=1, \\
			&0,\ \ \ \ \ \ \  r>1,
		\end{aligned}
		\right.
	\end{equation}
	indicating a short-range correlation.

	The other instance is a random state $\ket{\psi_{\text{Rand}}}$, 
	which is a typical volume-law entangled state, and can describe the equilibrium state of infinite-temperature chaotic systems.
	For $\ket{\psi_{\text{Rand}}}$, $\mathcal{E}_{\Psi, AB}$ is nearly independent of the positions of $A$ and $B$.
	Thus, the BPE of two distant qubits should be equivalent to the case of two nearest-neighbor qubits.
	According to Refs.~\cite{choi2023preparing,PhysRevLett.128.060601,PRXQuantum.4.010311}, $\mathcal{E}_{\Psi, AB}$ is nearly a Haar ensemble
	\begin{align} \label{Ehaar}
		\mathcal{E}_{\Psi, AB} \approx \mathcal{E}_{\text{Haar}, AB} =\{d\psi,\ket{\psi}\in \mathcal{H}_{AB}\},
	\end{align}	
	where $\mathcal{H}_{AB}$ is the Hilbert space of the qubits $A$ and $B$.
	Thus, we have 
	\begin{align} 
		\bar E(r) \approx \frac{\ln 2}{2},
	\end{align}	
	for arbitrary $r$, showing a long-range correlation.

\section{Ensemble averaged entanglements for stabilizer codes}\label{app3}
Here we present details about how to calculate the EAEs
of stabilizer states.
Consider a codeword $\ket{\psi}$ determined by the stabilizer group
\begin{align} 
	\mathcal{S} = \{\hat g_1,\hat g_2,...,\hat g_L: \hat g_\alpha\ket{\psi}=\ket{\psi}\},
\end{align}	
where $L$ is the number of qubits, $\hat g_\alpha$ is a Pauli string operator satisfying $[\hat g_\alpha,\hat g_\beta]=0$.
Now, for the wave function $\ket{\psi}$, we perform a projected measurement of the $j$-th qubit onto the $z$ component.

First, we consider a simple case, where $\ket{\psi}$ is the eigenstate of $\hat \sigma^z_j$, 
i.e.,$[\hat g_\alpha,\hat \sigma^z_j]=0 $ for all $\alpha$.
Thus, the measurement result is certain, and the wave function remains invariant after the projected measurement.

Second, we consider the case when $\ket{\psi}$ is not an eigenstate of $\hat \sigma^z_j$.
Here, without loss of generality, we consider $[\hat g_1,\hat \sigma^z_j]_+=0 $, 
while $[\hat g_\alpha,\hat \sigma^z_j]=0 $ for $\alpha\neq1$.
This can always be satisfied.
For instance, if there exists $\hat g_\beta$ ($\beta\neq1$) that satisfies $[\hat g_\beta,\hat \sigma^z_j]_+=0 $,
we can rewrite $\hat g_\beta$ as 
\begin{align} 
	\hat g_\beta\ \mapsto \hat g_1\hat g_\beta,
\end{align}	
and the new $\hat g_\beta$ commutes with  $\hat \sigma^z_j$.
Now the measurement result can be either $+1$ or $-1$ with equal probabilities.
If the measurement result is $+1$,  the wave function collapses to
\begin{align} 
	\ket{\psi} \mapsto \ket{\tilde{\psi}_+} = \frac{1+\sigma^z_j}{2}\ket{\psi}.
\end{align}	
We can also find that $\ket{\tilde{\psi}}$ satisfies
\begin{align} \nonumber
	& \hat g_\alpha \ket{\tilde{\psi}_+} = \ket{\tilde{\psi}_+} \ \ \ (\alpha\neq1),\\
	&\sigma^z_j \ket{\tilde{\psi}_+} = \ket{\tilde{\psi}_+} .
\end{align}	
Thus, the new codeword $\ket{\tilde{\psi}_+}$ is determined by the following stabilizer group
\begin{align} 
	\tilde{\mathcal{S}}_+ = \{\hat g_1=\hat \sigma^z_j ,\hat g_2,...,\hat g_L: \hat g_\alpha\ket{\tilde{\psi}_+}=\ket{\tilde{\psi}_+},
\end{align}		
Similarly, if the measurement result is $-1$, then the wave function after the projected measurement, labeled by $\ket{\tilde{\psi}_-} $ ,
is determined by the following stabilizer group
\begin{align} 
	\tilde{\mathcal{S}}_- = \{\hat g_1=-\hat \sigma^z_j ,\hat g_2,...,\hat g_L: \hat g_\alpha\ket{\tilde{\psi}_-}=\ket{\tilde{\psi}_-}.
\end{align}		
Thus, the difference between $\tilde{\mathcal{S}}_+$ and $\tilde{\mathcal{S}}_-$ is just that the corresponding $\hat g_1$ has the opposite sign.

Note that the entanglement entropy of a codeword $\ket{\psi}$ is independent of the signs of $\hat g_\alpha$. 
Thus, $\tilde{\mathcal{S}}_+$ and $\tilde{\mathcal{S}}_-$ have the same entanglement entropies.
Generalizing to the case of joint projected measurements of many qubits,
we can conclude that the entanglement entropy of a codeword after the measurements is independent of the measurement results.
Therefore, for a codeword $\ket{\psi}$, we consider the BPE with respect to the $z$-component basis  as
\begin{align} 
	\mathcal{E}_{\psi, AB} := \{p(z_R),\ket{\Psi_{AB}(z_R)}\}.
\end{align}	
The  von Neuman entropy of subsystem $A/B$ for the state $\ket{\Psi_{AB}(z_R)} $ is $z_R$-independent.
Therefore, when calculating the EAE of $\mathcal{E}_{\psi, AB}$, we can first perform a projected measurement of the subsystem $R$.
Then calculate the von Neuman entropy of subsystem $A/B$, which is the EAE of $\mathcal{E}_{\psi, AB}$.

\section{Haar random quantum circuits with projective measurements}\label{app1}
\begin{figure*}[t] \includegraphics[width=0.9\textwidth]{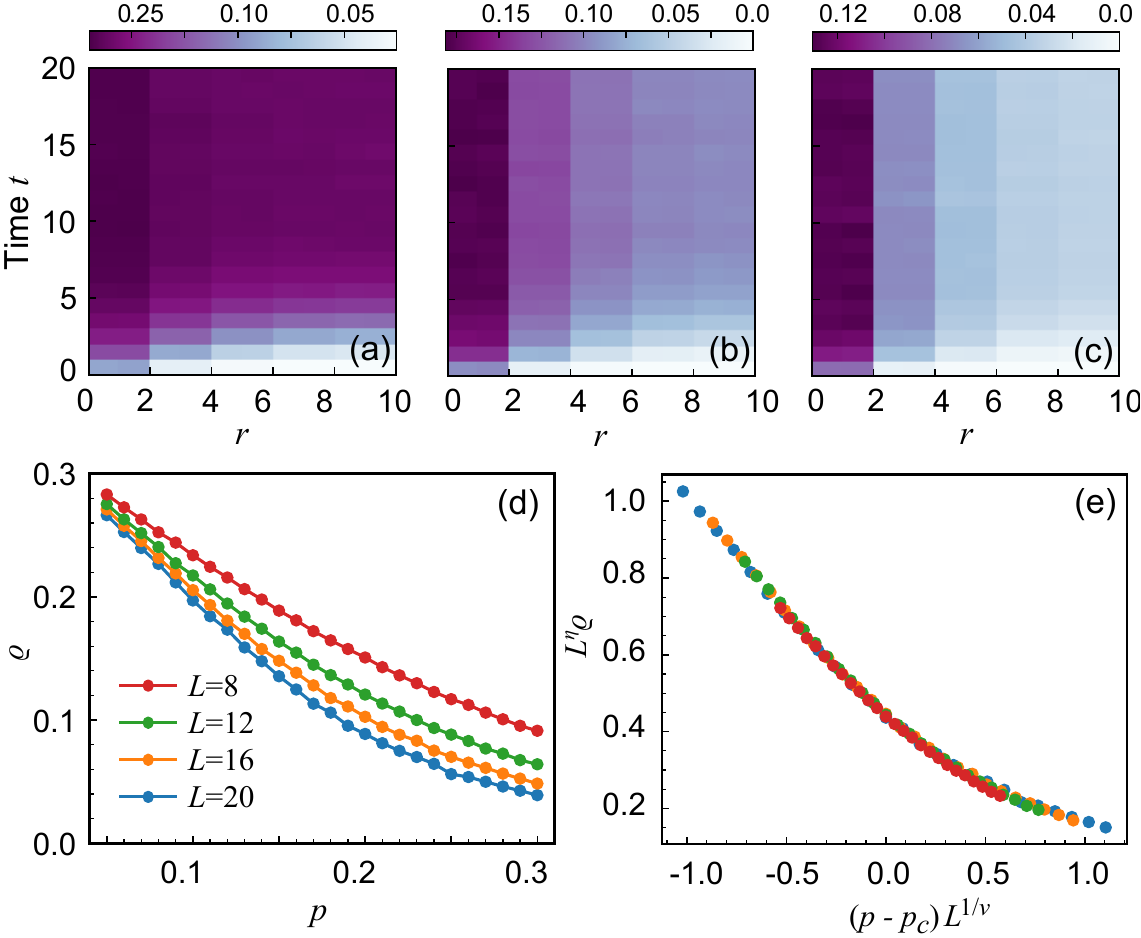}
	\caption{The results of EAE for Haar random circuits.
		The dynamics of $\bar E(r)$ for (a) $p=0.05$, (b) $p=0.17$, and (c) $p=0.25$. The system size is $L=20$.
		(d) The integrated EAE $\varrho$ of steady states versus $p$ for different system sizes. 
		(e) Data collapse of (d) using Eq.~(6) with the critical point $p_c \approx 0.17$ and exponents $\nu\approx1.4$ and $\eta \approx0.45$.}
	\label{fig_a1}
\end{figure*}

In the main text, we have applied BPE and EAE to investigate MIPT in the measured stabilizer circuits.
In this section, we apply BPE and EAE to study MIPT in Haar random quantum circuits with projective measurements.
In a Haar random quantum circuit, each two-qubit gate is sampled according to the Haar measure on the unitary group  $SU(4)$.
The numerical results are shown in Fig.~\ref{fig_a1} with periodic boundary conditions.
We find that EAE can indeed identify the MIPT of Haar random quantum circuits with projective measurements, 
and the critical point is consistent with Ref.~\cite{PhysRevB.101.060301}.
We also calculate the critical exponents as $\nu\approx1.4$ and $\eta \approx0.45$.

\section{Finite-size scaling}\label{app4}

In the main text, we have presented the universal scaling ansatz and the corresponding numerical results of the EAE across the critical point.
In this appendix, we derive the scaling functions of the EAE across a generic MIPT.
Our derivations are all premised on a hypothesis: the role of EAE in MIPTs is equivalent to the correlation function in conventional phase transitions.

\subsection{Static scaling ansatz}
We first consider the scaling ansatz of steady states in monitored quantum systems.
Near a critical point, the correlation length $\xi$  for a steady state diverges as
\begin{align} \label{xi}
	\xi \sim (p-p_c) ^{-\nu}.
\end{align}	
%
Thus according to scaling invariance, an arbitrary observable $\mathcal{O}$ can be described by a universal scaling function
\begin{align} 
	\mathcal{O}  = L^{-\Delta_{\mathcal{O} }}f\big(\frac{\xi}{L}\big),
\end{align}	
where $\Delta_{\mathcal{O} }$ is the dimension of $\mathcal{O}$.
According to Eq.~(\ref{xi}), we have 
\begin{align} 
	\mathcal{O} \  =\  L^{-\Delta_{\mathcal{O} }}f\big[\frac{(p-p_c) ^{-\nu}}{L}\big] \ =\  L^{-\Delta_{\mathcal{O} }}F\big[(p-p_c)L^{1/\nu}\big],
\end{align}	
where the function $F(x)=f(x^{-1/\nu})$.
For the integrated EAE $\varrho := \frac{1}{L}\sum_{r=1}^{L}\bar E(r)$, 
since $\bar E(r)\sim r^{-\eta}$ at the critical point,
we have
\begin{align} 
	\varrho \sim \frac{1}{L}\int_0^L dr \ r^{-\eta}  \sim L^{-\eta}.
\end{align}	
Thus, the dimension of $\varrho $ is $\Delta_\varrho  = \eta$, and it should obey the scaling ansatz
\begin{align} 
	\varrho = L^{-\eta}F[(p-p_c)L^{1/\nu}].
\end{align}	

\begin{figure*}[t] \includegraphics[width=1\textwidth]{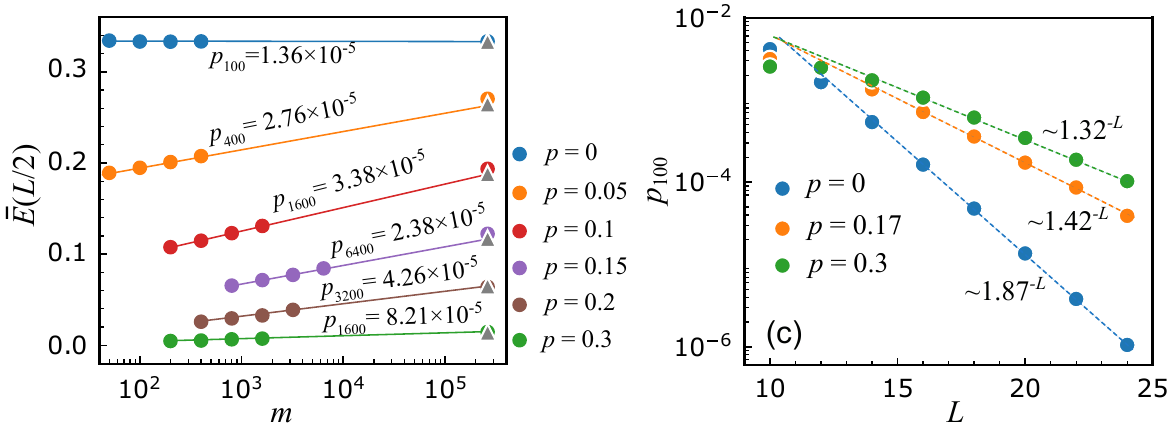}
	\caption{ Complexity of probing EAE in monitored random Haar circuits. (a) Extrapolation of $\bar E(L/2)$ with $L=20$ for different monitoring rates $p$.
		The solid line is the fit according to Eq.~(\ref{E_fit}), and the gray triangles shows the extrapolation values.
		The complexities  of probing EAE, i.e., $p_m$, are also presented.
		(b) The scaling of $p_{100}$ for different monitoring rates $p$.}
	\label{fig_a2}
\end{figure*} 

\subsection{Dynamical scaling ansatz}
Since MIPT is a nonequilibrium phase transition,
we can also study  dynamical scaling.
We first consider the thermodynamic limit, i.e., $L\rightarrow \infty$.
In addition to the correlation length $\xi$ of the steady sate, there also exists a dynamical correlation length $\xi (t)$.
Here, $\xi (t)$ can be understood as the  instantaneous correlation length at time $t$, and satisfies
\begin{align} \label{xit}
	\xi (t) = t^{1/z},
\end{align}	
where $z$ is the dynamical critical exponent.
Similar to the dynamical scaling of conventional phase transitions~\cite{Altland2010}, 
the dynamics of EAE is expected to satisfy the scaling ansatz
\begin{align} \label{Erta} \nonumber
	\bar E(r,t,p) &= t^{\theta-d/z} h[r/\xi(t),\xi(t)/\xi] \\ \nonumber
	&= t^{\theta-d/z} h[r/t^{1/z},(p-p_c)^{\nu}t^{1/z}] \\  
	&= t^{\theta-d/z} g[r/t^{1/z},(p-p_c)t^{1/\nu z}],
\end{align}	
where $d=1$ is the spatial dimension,  $\theta$ is another universal critical exponent, 
and the scaling functions satisfy $g(x,y)=h(x,y^{1/\nu})$.
Thus, the $k$-moment of EAE 
satisfies
\begin{align} \label{rhokt_sa}\nonumber
	\varrho^{(k)}(t,p) & \sim \frac{1}{L^{k+1}}\int_0^L dr\ r^{k}t^{\theta-d/z} g[r/t^{1/z},(p-p_c)t^{1/\nu z}] \\
	& \sim t^{\theta+k/z} G_k[(p-p_c)t^{1/\nu z}],
\end{align}
where
\begin{align}\nonumber
	G_k (y) = \int dx\ g(x,y).
\end{align}
At the critical point $p=p_c$, according to Eq.~(\ref{rhokt_sa}), we have 
\begin{align} 
	\varrho^{(k)}(t,p_c)  \sim t^{\theta+k/z} G_k(0),
\end{align}
showing that $\varrho^{(k)}$ exhibits a power-law increase at the critical point.

Now we consider finite-size systems, where the correlation length of a steady state satisfies $\xi = L$ at the critical point.
Therefore, according to Eq.~(\ref{Erta}), we have
\begin{align}  \label{Ert_fs} \nonumber
	\bar E(r,t,p_c) &= t^{\theta-d/z} h[r/\xi(t),\xi(t)/\xi] \\ \nonumber
	&= t^{\theta-d/z} h[r/t^{1/z},t^{1/z}/L] \\  
	&= t^{\theta-d/z} G[r/t^{1/z},tL^{-z}],
\end{align}	
where $G(x,y) = h(x,y^z)$.
For the $k$-moment of EAE $\varrho^{(k)}$, similar to Eq.~(\ref{rhokt_s}), integrating $r$, we can obtain
\begin{align}\label{rho_tL}
	\varrho^{(k)}(t,p_c)   \sim t^{\theta+k/z} G_k(tL^{-z}).
\end{align}
Here, we note that Eq.~(\ref{Ert_fs}) is only valid when  $t\ll L$.
Therefore, the scaling ansatz Eq.~(\ref{rho_tL}) is also only valid for early times.

	\section{Complexity of probing EAE in quantum simulators}\label{app5}

	In this appendix, we discuss the complexity of measuring EAE in quantum simulators with more details,
	where we mainly choose monitored Haar random circuits as an example.
	
	According to  Eq.~(\ref{eae}) in the main text, to measure $\bar E(A:B)$, 
	we only need to measure single-qubit density matrices of $A$,
		\begin{align} 
		\hat \rho_A (z_R):=\text{Tr}_B\ket{\Psi_{AB}(z_R)}\bra{\Psi_{AB}(z_R)}.
	\end{align}
	In quantum-simulation experiments, it is challenging to obtain all states $\ket{\Psi_{AB}(z_R)}$ in the ensemble $	\mathcal{E}_{\Psi, AB}$.
	Instead, we can obtain the partial density matrices $\hat \rho_A (z_R)$ with the largest probabilities $p(z_R)$, 
	where we label the number of density matrices as $m$.
	Then, we can approximate EAE with these $m$ density matrices as 
	\begin{align} 
		\bar E_m(A:B) = \frac{1}{\sum _{j=1}^m p(z_R^{(j)})}\sum _{j=1}^m p(z_R^{(j)})S[\hat \rho_A (z_R^{(j)})],
	\end{align}
	where $z_R^{(j)}$ ($j=1,2,...,m$) is the corresponding measurement results of $R$.
	According to Fig.~\ref{fig_a2}(a), we can find that $\bar E_m(A:B)$ satisfies
	\begin{align} \label{E_fit}
		\bar E_m(A:B) = \bar E_0 + A\log m. 
	\end{align}
	Thus, by using the extrapolation, we can obtain $\bar E(A:B)$ for the full ensemble $\mathcal{E}_{\Psi, AB}$.
	In Fig.~\ref{fig_a2}(a), we can find that the extrapolation results of $\bar E(A:B)$ are close to the true values (i.e., $m=2^{L-2}$).

	Now we discuss the complexity of measuring EAE in quantum simulators, i.e., how many repetitive readouts are necessary.
	The number of measurements depends on the $m$th largest probability $p(z_R)$, labeled by $p_m$.
	For a many-qubit system, intuitively, $p_m$ decays exponentially when increasing the system size , i.e, 
	\begin{align} 
		p_m = q^{-L}.
	\end{align}
	Thus, the complexity of measuring EAE scales as the inverse of $p_m$, i.e, $q^L$.
	Here, since $m\ll\mathcal{D}$, where $\mathcal{D}$ is the dimension of the Hilbert space,  $p_m$ should satisfy
	\begin{align} 
		p_m > \frac{1}{\mathcal{D}}.
	\end{align}
	Thus, we have 
	\begin{align} 
		q\leq2.
	\end{align}
	In Fig.~\ref{fig_a2}(b), we present the 100th largest $p(z_R)$ for the monitored random Haar circuits.
	We can find that $p_{100}$ exhibits exponential decay when increasing the system size.
	In addition, the corresponding $q$ becomes small when increasing the monitoring rate (the entanglement entropies become small).

	According to Fig.~\ref{fig_a2}(b), for a larger $p$, i.e., the lower-entangled system, although it needs larger $m$ to fit  EAE,
	$p_m$ is exponentially larger than the one of $p=0$.
	Thus, the corresponding complexity should not be larger than the case of $p=0$.
	According to Fig.~\ref{fig_a2}(a), we can find that, to make EAE accurate enough, the case of $p=0$ has the largest complexity.
	Specifically, when $p=0$ and $L=24$ (i.e., the Haar random circuits without monitoring), 
	$p_{100}\approx10^{-6}$.
	Figure~\ref{fig_a2}(a) shows that $m=100$ is enough to obtain an accurate EAE,
	so we need about $10^6$ repetitive readouts.
	For superconducting qubits, $10^6$ single-shot readouts can be accomplished in about one hour.
	However, for a 24-qubit system, measuring the half-chain entanglement entropies needs about  $10^8$ times of readouts~\cite{google2023measurement}.
	Therefore, measuring EAE is easier than measuring entanglement entropies in quantum simulators.

\end{appendix}

%


\begin{thebibliography}{74}
\bibitem [{squ()}]{squareM}%
	\BibitemOpen
	\bibinfo {title} { The squared magnetization is $M_2 : =
			(\frac{1}{L}\sum_j \sigma_j)^2$, where $\sigma_j = \pm1$ is a classical spin.
			$M_2$ contains arbitrary two-site correlation functions. Since the EAE can be
			considered as a correlation function, $\varrho$ has a similar meaning as
			$M_2$ in the Ising model}\ \href@noop {} \BibitemShut {NoStop}%
\end{thebibliography}

\end{document}